\documentclass[reprint,prl,aps,floatfix,unsortedaddress]{revtex4-1}
\usepackage{graphicx}
\usepackage{amsmath,amsfonts,mathcomp}
\usepackage[modulo]{lineno}
\usepackage{siunitx}
\usepackage{xcolor,colortbl,tabularx,multirow}
\usepackage{listings}
\usepackage{enumerate}
\usepackage[T1]{fontenc}      
\usepackage{array}

\newcolumntype{L}[1]{>{\raggedright\let\newline\\\arraybackslash\hspace{0pt}}m{#1}}
\newcolumntype{C}[1]{>{\centering\let\newline\\\arraybackslash\hspace{0pt}}m{#1}}
\newcolumntype{R}[1]{>{\raggedleft\let\newline\\\arraybackslash\hspace{0pt}}m{#1}}

\begin{document}  
\title{Effect of a Tertiary Butyl Group on Polar Solvation Dynamics in Aqueous Solution: A Computational Approach}
\author{Esther Heid and Christian Schr\"oder}
\email{christian.schroeder@univie.ac.at}
\affiliation{University of Vienna, Faculty of Chemistry, Department of Computational Biological Chemistry, 
W\"ahringerstra{\ss}e 19, A-1090 Vienna, Austria}

\begin{abstract}
The current computational study investigates the changes in solvation dynamics of water when introducing hydrophobic side chains to the molecular probe N-methyl-6-oxyquinolinium betaine.
High precision transient fluorescence and absorption measurements published in the companion paper revealed an influence of hydrophobic sidechain alterations
on the observed solvation dynamics of a chromophore in water.
As the influence of shape, size and structure of chromophores on the time-dependent Stokes shift was so far
thought to play a role only in slowly rotating solvents compared to the solute or if the hydrogen bonding ability of the solute changes, this finding is quite unexpected.
Analysis of the time-dependent Stokes shift obtained from nonequilibrium simulations corroborates experimental retardation factors and activation energies, and indicates that
solute motion, namely vibration, is mainly responsible for the observed retardation of solvation dynamics. The faster dynamics around the smaller chromophore is in fact achieved by some normal modes located
at the pyridinium part of the chromophore. Rotation also contributes to a very small extent to hydration dynamics, but for 
small and large derivatives alike.
Local residence times furthermore reveal slight retardations  in the first solvent shell around the chromophores.
The current picture of the solute acting as a passive
molecular probe therefore  needs to be revised even for solvents like water.
\end{abstract}
\maketitle

\section{Introduction}
Time-dependent fluorescence deals with the timescale of solvent rearrangement following an electric perturbation which is realized via an optical excitation of a suitable chromophore. Numerous chromophores 
are available, and current research indicates that the choice of chromophore may affect the outcome of such an experiment. Although the observed dynamics are largely
 a measure of solvent properties, the structure~\cite{sim89a,bar91a,lad94a,mar95c} and charge redistribution~\cite{sch17a,mar95b,mar98a} of the solute have been shown to play a role, too. In slowly rotating solvents such as alcohols, the size of the chromophore can  
 influence the observed solvent relaxation function, as for example shown by Su and Simon who observed an about 15\% faster solvation of dimethylbenzonitrile compared to diethylbenzonitrile in propanol~\cite{sim89a}. 
 Analogously, Barbara and coworkers found that various solvents relax 20-40\% faster around the smaller Coumarin 152 than around Coumarin 153~\cite{bar91a}. Contrary results were simulated for diatomic 
 artificial solutes differing only in size, where a faster rearrangement was found for the larger solute~\cite{lad94a}, indicating that the above described outcomes may not purely be an effect of solute 
 volume. However, the study was conducted via equilibrium molecular dynamics simulations and invocation of linear response theory using the solvent methanol, for which it was shown that linear response 
 approximations are likely to fail~\cite{sch17a}. 
 Effects of solute motion were also observed for small benzene-like solutes in methanol and acetonitrile~\cite{mar95b}. Extrapolation of these results to large, real chromophores, however, predicted no significant contribution of solute motion, at least
 regarding equilibrium simulations making use of the linear response theory. 
 
 An important step towards fuller understanding of the size dependence of solvation dynamics was made by
 Ernsting and coworkers who correlated solvation dynamics in various solvents with the rotational diffusion time of the solute~\cite{ern11a}. 
 Solute dependence of solvation dynamics was found in slowly rotating methanol, but not in water. The authors concluded that
whenever the solute rotates on a similar timescale as the solvent, the solute rotational 
 contribution to the time-dependent Stokes shift is non-negligible (see also Ref.~\citenum{mar95b}). For solvents like water, this type of molecularity escapes detection.
It should also be mentioned that the  structure of the chromophore, namely its ability to specifically  interact with the solvent via hydrogen bonds can also affect solvation dynamics, as found by Maroncelli 
and coworkers by measuring the solvation response of diverse solutes in propanol~\cite{mar95c}.

 A solute dependence of solvation dynamics in aqueous solution, with water rotating faster than all solutes, should therefore only be of importance for solutes differing in their hydrophilic sites, 
 where hydrogen-bonding may occur, or for varying charge redistributions upon excitation. However, we find that various derivatives of N-methyl-6-oxyquinolinium betaine (MQ), differing only in their 
 hydrophobic side chains and not in their hydrophilic site, give not the same  solvation response in water. The Stokes shift of MQ itself has been well investigated, 
 both by experiment~\cite{ern05a,ern10a,ern11a} and simulation~\cite{seb13a,reg14a,sch16b}, but the exploration of side chain effects has not been described before.
 High resolution time-dependent fluorescence measurements, published in the companion paper, Ref.~\citenum{ern17a},
 of MQ and a derivative containing a tertiary butyl group revealed an interesting influence of hydrophobic side chain substitutions.
 The solvation dynamics around the chromophore slowed down
upon introduction of the bulky alkyl side chain, revealing a slight retardation of water dynamics by the hydrophobic moiety. The present simulation study 
 analyzes the experimental results via  nonequilibrium molecular dynamics simulation and aims to predict the behavior of seven derivatives of MQ in aqueous solution. 
The atomistic simulation furthermore enables the 
identification of the character and locality of the effect, and reveals contributions of solute motion, especially solute vibration.
The use of nonequilibrium simulations proved to be especially important, as equilibrium simulations fail to describe retardation of the larger chromophore and the molecularity would escape detection even in simulation.
The combination of experiment and simulation therefore enables the detailed analysis of the molecularity of aqueous solvation dynamics both on a qualitative and quantitative scale, and contributes significantly to the 
understanding of time-dependent fluorescence in quickly relaxating solvents.

\section{Methods}
\begin{figure}[t]
 \includegraphics[width=0.7\linewidth]{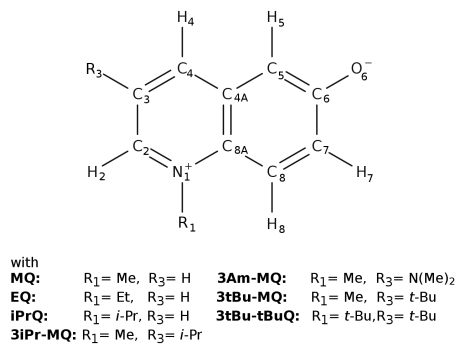}
\caption{Derivatives of 6-oxyquinolinium betaine used in this study.}
\label{FIG:derivatives}
\end{figure}

All simulations were conducted with the program package CHARMM~\cite{kar09a}. The flexible, atomistic force field for N-methyl-6-oxyquinolinium betaine (MQ) was taken from Ref.~\citenum{sch16b}, where we compared the influence
of different functionals, basis sets and charge assignment methods, as well as water force fields on the time-dependent Stokes shift. 
The force fields of the derivatives
of MQ, as shown in Fig.~\ref{FIG:derivatives}, were calculated accordingly using the  method that produced results in closest agreement to experiment.
Namely, the solutes were optimized at the B3LYP/6-311G++(2d,2p) level of theory and the partial charges calculated with DFT/hybrid DFT for the ground state 
and TD-DFT for excited states in Gaussian 09~\cite{gau09a}. We used the $\omega$B97xD hybrid DFT functional~\cite{cha08a}, an aug-cc-pVTZ basis set, implicit solvation via the polarizable continuum model (PCM) of water~\cite{tom05b}
and the CHelpG~\cite{wib90a} method to assign partial charges.
All other force field parameters were obtained via PARAMCHEM~\cite{van12a,van12b} which is based on the CHARMM General Force Field (CGenFF)~\cite{van10a}. Forcefields for all employed chromophores can be found in the Supporting Information.
According to the results of Ref.~\citenum{sch16b}, we chose to use the polarizable SWM4-NDP water model~\cite{rou03b}, as this was shown to reproduce experimental solvation dynamics most accurately.

Bonds containing hydrogens were kept at constant length with the SHAKE algorithm.
The solute was allowed to rotate, translate and vibrate freely, unless stated otherwise.
Simulations were carried out in cubic boxes which were obtained by randomly putting one solute molecule and 1000 molecules of  water  in a box of sidelength \SI{32}{\angstrom} with PACKMOL~\cite{mar09a}
and equilibrating each system at the temperature of interest (\SI{4}{\celsius}, \SI{8}{\celsius}, \SI{20}{\celsius}, \SI{27}{\celsius}, \SI{37}{\celsius} and \SI{60}{\celsius}) in an NPT ensemble. The box-lengths converged to the values given in Table~\ref{TAB:boxl}.
\begin{table}[t]
 \centering
 \caption{Converged box-lengths of the simulated systems}
 \begin{tabular}{lcccccc}
 \hline\hline\
  &\multicolumn{6}{c}{Boxlength [\AA]}\\
 &\SI{4}{\celsius}&\SI{8}{\celsius}&\SI{20}{\celsius}&\SI{27}{\celsius}&\SI{37}{\celsius}&\SI{60}{\celsius}\\ 
\hline
 MQ 	&31.04&31.07&31.14&31.20&31.27&31.48\\
 EQ&-&-&-&31.20&-&-\\
 iPrQ&-&-&-&31.21&-&-\\
 3iPr-MQ&-&-&-&31.22&-&-\\
 3Am-MQ&-&-&-&31.22&-&-\\
 3tBu-MQ &31.06&31.09&31.18&31.22&31.31&31.52\\
 3tBu-tBuQ&-&-&-&31.24&-&-\\
   \hline\hline\\
 \end{tabular}
 \label{TAB:boxl} 
\end{table}

A Nos\'e-Hoover thermostat~\cite{nos84a,hoo85a} was used to control temperature. Periodic boundary conditions were applied, where electrostatic interactions were calculated via the Particle Mesh Ewald method with grid size of about \SI{1}{\angstrom},
cubic splines of order 6 and a $\kappa$ of \SI{0.41}{\angstrom}$^{-1}$. Van der Waals interactions were cut off at \SI{11}{\angstrom}.

For the calculation of equilibrium rotation times and Stokes shifts from linear response theory a \SI{10}{ns} NVT trajectory at \SI{27}{\celsius} (after an equilibration of \SI{1}{ns}) was written. For the calculation of the nonequilibrium 
Stokes shift relaxation function at least 500 independent starting configurations per system
were obtained from a long NVT run at elevated temperature. Whenever small differences between systems were evaluated, the number of configurations was raised up to 2000.
Each configuration was then equilibrated for \SI{500}{\pico\second}, 
afterwards the partial charge distribution was changed to the excited state, and the following \SI{50}{ps} were monitored, where the trajectory
was saved in logarithmic intervals to save disk space. The excitation was therefore only described by a change in charge, and the remaining force field and geometry was left unchanged. However, a test simulation allowing geometry changes according to quantum mechanical 
geometries did not show a significant influence on the Stokes shift.

The electrostatic contribution to the band gap correlation function, in the following termed the Stokes shift relaxation function $S(t)$,
\begin{equation}
 S(t)=\frac{\overline{\Delta U(t)} - \overline{\Delta U(\infty)}}{\overline{\Delta U(0)} -\overline{\Delta U(\infty)}}
 \label{EQU:S}
\end{equation}
where $\overline{\Delta U(t)}$ is the average interaction energy between solute and solvent, was calculated using a Python program based on MDAnalysis~\cite{bec11a}. To calculate the integral solvation time
\begin{equation}
 \tau_\mathrm{tot}=\int_0^\infty S(t) \mathrm{d}t
 \label{EQU:S00}
\end{equation}
$S(t)$ was fitted to a sum of a Gaussian and a Kohlrausch-William-Watts function, so that $\tau_\mathrm{tot}$ could be obtained by analytically integrating the fitting function. 
The preceding experimental work~\cite{ern17a} suggested to fit the relaxation function with up to three exponential decay functions solely after \SI{300}{fs}, where 
only diffusive terms contribute to the relaxation, and 
extrapolate to $t=0$, to obtain a full description of this overdamped motion. Throughout this work we therefore  give also the relaxation time of collective rotation $\tau_\mathrm{solv}$, where 
we used only a Kohlrausch-William-Watts function
and fitted $S(t)$ from \SI{0.3}{ps}. The use of a three exponential decay (similar to experiment) leads to large uncertainties in the fitting routine for simulation, as data is sampled in different time intervals and therefore 
not enough data points exist for each of the exponential
functions. As the fitting function is only used for integration and does not resemble physical processes, the results of this study are independent of the respective fit function.

\section{Results and discussion}
\subsection{Stokes shift relaxation function, site-specific residence times and equilibrium response}
\begin{figure}[t]
 \includegraphics[width=\linewidth]{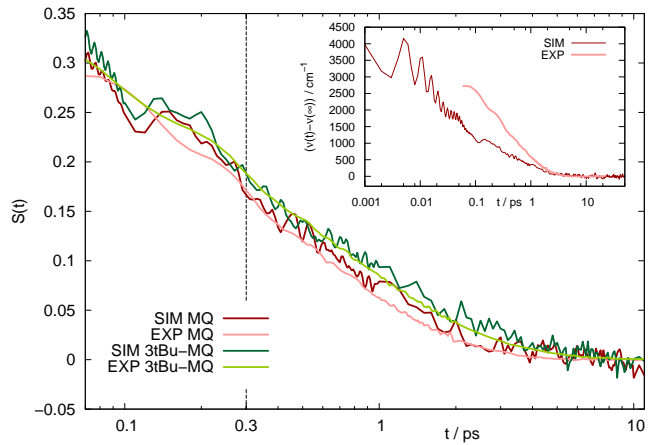}
\caption{Comparison of simulated data of MQ and 3tBu-MQ at \SI{20}{\celsius} to experiment~\cite{ern17a}, where experimental data was scaled so that the curves correspond to each other at \SI{0.3}{ps} (vertical dashed line). 
The inset shows the absolute, unnormalized shift of MQ from simulation and experiment.}
\label{FIG:exp}
\end{figure}

Figure~\ref{FIG:exp} shows the Stokes shift relaxation function  of the original solute MQ as well as of the largest derivative 3tBu-MQ obtained from nonequilibrium simulation and experiment at \SI{20}{\celsius}.
The experimental values have been
 scaled to match the data from simulation at t=\SI{0.3}{ps}.
The relaxation functions from both simulation and experiment show the same trend: The solvation dynamics of 3tBu-MQ is slower than for MQ, experimentally by a factor of about 1.4, in simulation by a factor of about 1.3. 
The shape of the respective simulated curves corresponds well to the experiment, reinforcing the correctness of the used solute and solvent models, as well as the high experimental accuracy. 
\begin{table}[t]
 \centering
 \caption{$\tau$ of MQ and 3tBu-MQ at different temperatures. Experimental values from the companion paper, Ref.~\citenum{ern17a}.}
 \begin{tabular}{lrccc}
 \hline\hline\\
			&&\multicolumn{2}{c}{SIM}&EXP\\
			&&$\tau_\mathrm{tot}$ {[ps]}&$\tau_\mathrm{solv}${[ps]}& $\tau_\mathrm{solv}$ {[ps]}\\\hline
\SI{4}{\celsius}&MQ  		&0.42	& 0.86	\\
		&3tBu-MQ  	&0.65   & 1.32  \\
\SI{8}{\celsius}&MQ  		&0.41   &0.82	 & \textcolor{black}{0.91}\\
		&3tBu-MQ  	&0.47	&0.96	&\textcolor{black}{1.40}\\
\SI{20}{\celsius}&MQ  		&0.30	&0.68	&\textcolor{black}{0.57}\\
		&3tBu-MQ 	&0.38	&0.92	 &\textcolor{black}{0.76}\\
\SI{27}{\celsius}&MQ  		&0.24	&0.56	 \\
		&3tBu-MQ	& 0.34	&0.80	 \\
\SI{37}{\celsius}&MQ 		&0.24	&0.49	&\textcolor{black}{0.43}\\
		 &3tBu-MQ 	&0.25	&0.57	&\textcolor{black}{0.53}\\
\SI{60}{\celsius}&MQ 		&0.16	&0.39	\\
		  &3tBu-MQ 	&0.15	&0.36	\\
 \hline\hline\\\
 \end{tabular}
 \label{TAB:tautemp} 
\end{table}
The inset in Fig.~\ref{FIG:exp} shows the absolute shifts of experiment and simulation.
Although the normalized relaxation functions fit well as explained above, the absolute functions show deviations, which  consist mainly of a constant scaling factor. 
This difference could arise from the approximations made in simulation which are outlined in the Methods section. For example, the solute can vibrate freely but the change in internal energy of the chromophore is not taken into account in simulation, 
\textit{i.e.} is not included in $\overline{\Delta U(t)}$ in Eq.~\ref{EQU:S}. Also,
no changes in force constants were made upon excitation. 
Furthermore we expect that some differences  can be  overcome by 
inclusion of solute polarizability, which is known to slow down solvation dynamics~\cite{kim95a,kim05a,mar95b}. This influence of polarizability will be investigated in a future publication.
However, as this systematic error occurs for all chromophores in this study alike and consists only of a factor, the timescale of relaxation and thus the difference in solvation dynamics between MQ and 3tBu-MQ is still
expected to be described correctly, reflecting the timescales seen in experiment. 

\begin{figure}[t]
 \includegraphics[width=\linewidth]{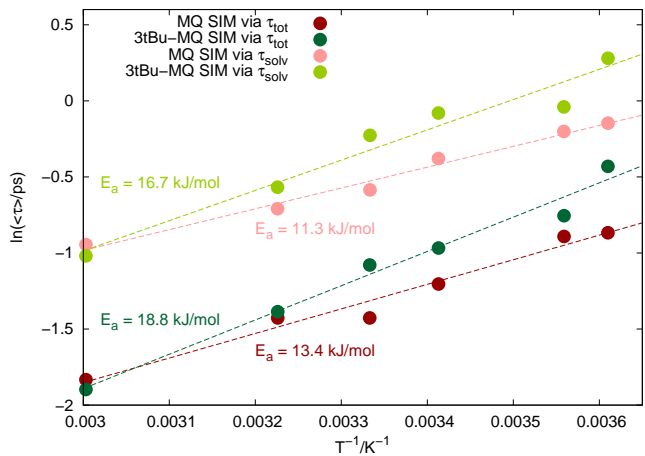}
\caption{Arrhenius plots of the solvation time $\tau_\mathrm{tot}$ and $\tau_\mathrm{solv}$. An activation enthalpy of \SI{13.4}{kJ/mol} for MQ and \SI{18.8}{kJ/mol} for 3tBu-MQ is observed for the total relaxation and
\SI{11.3}{kJ/mol} for MQ and \SI{16.7}{kJ/mol} for 3tBu-MQ for collective rotation.}
\label{FIG:arrh}
\end{figure}

Relaxation times obtained via Eq.~\eqref{EQU:S00} for MQ and 3tBu-MQ are listed in Table~\ref{TAB:tautemp} at different temperatures and also fit very well to  experiment
throughout the observed temperature range.
The temperature dependence of the 
relaxation time can be used to calculate an activation energy of rearrangement, where both $\tau_\mathrm{tot}$ and  $\tau_\mathrm{solv}$ were used respectively. 
Different activation energies were found for MQ and 3tBu-MQ, as indicated 
by the different slopes of the Arrhenius plot (ln$\tau_\mathrm{tot}$ plotted
against 1/$T$) shown in Fig.~\ref{FIG:arrh}. 
The observed activation enthalpy for MQ of \SI{13.4}{kJ/mol}      
is smaller than the value for 3tBu-MQ of \SI{18.8}{kJ/mol}.
For the correlated rotation relaxation after \SI{300}{fs} an activation enthalpy of \SI{11.3}{kJ/mol} and \SI{16.7}{kJ/mol} is obtained. Despite the difference in absolute values, the retardation corresponds to
about \SI{5.4}{kJ/mol} in both cases, in agreement to experimental results
of the companion paper, Ref.~\citenum{ern17a}, where an increase in activation energy of \SI{5.0}{kJ/mol} was found.

 To rule out the possibility of changed solvation dynamics solely via an inductive effect of the sidechain we also conducted simulations where for 3tBu-MQ either the change in charge $\Delta q$ or both the ground state charge $q_{\mathbb{S}0}$ and $\Delta q$ were
set to the corresponding values encountered in the unsubstituted derivative MQ. However, these
artificial changes in charges did not reduce the retardation observed, but rather strengthened it. 
Furthermore we monitored the influence of geometry adaptations caused by sidechain addition, but found no influence on the time-dependent Stokes shift. The respective data for both simulations is shown in the Supporting Information.

Additionally, we conducted equilibrium simulations of the ground state of the chromophores in water, as
Kumar and Maroncelli found solute dependence in equilibrium simulations of solvation dynamics for small solutes like benzene, where solute motion contributes significantly to the dynamics~\cite{mar95b}. 
They concluded that for larger chromophores the effects of probe motion should 
be negligible, at least for systems where linear response theory is valid. 
The solvation response function
via the linear response approximation is
\begin{equation}
  S(t) \simeq \frac{\langle\delta\Delta U(0)\delta\Delta U(t)\rangle}{\langle\delta\Delta U(0)^2\rangle} = C(t)
\end{equation}
where $\delta\Delta U$ is defined as
  \begin{equation}
 \delta\Delta U(t) = \Delta U(t) - \langle\Delta U \rangle
\end{equation}
and $\langle\ldots\rangle$ denotes a mean value. Interestingly, the solute dependence of the solvation response vanishes almost completely for $C(t)$ as predicted in 
Ref.~\citenum{mar95b} and the distinct differences between MQ and 3tBu-MQ  seen in $S(t)$ is lost (results shown in the Supporting Information).
When conducting equilibrium simulations in the excited state, however, a very small difference in $C(t)$ between MQ and 3tBu-MQ is observed.  The difference in MQ and 3tBu-MQ is therefore not a static effect
but a dynamic and possibly even nonlinear effect.
However, we note that $C(t)$ is a poor description of $S(t)$ for the system under investigation and gives quantitatively wrong results, so that comparison of $C(t)$ for different chromophores is not applicable.
More detailed information on the applicability of linear response theory in this system is given in the Supporting Information.

To further examine
the curious finding of different relaxation times and activation enthalpies for chromophores that have the same chemical structure concerning their polar regions, different derivatives of oxyquinolinium betaine were simulated at \SI{27}{\celsius} 
(via nonequilibrium simulations). 
The corresponding relaxation times are listed in Table~\ref{TAB:taures}. A clear trend can be seen, namely 
that longer hydrophobic side chains cause longer relaxation times, consistent with the observations of MQ and 3tBu-MQ. EQ shows slightly longer relaxation times as expected, which is attributed to the uncertainty of the simulation and fitting routine.
\begin{table}[t]
 \centering
 \caption{Simulated relaxation time of different derivatives (see Fig.~\ref{FIG:derivatives}) at \SI{27}{\celsius}, as well as rotational diffusion timescales  obtained from the second Legendre polynomial of the normalized
correlation function of the solute dipole vector (for comparison, the rotation time of the SWM4 water model is approximately \SI{4}{ps}). }
 \begin{tabular}{lC{1.5cm}C{1.5cm}C{1.5cm}}
 \hline\hline\
	&$\tau_\mathrm{tot}$ {[ps]} 	&$\tau_\mathrm{solv}${[ps]}&	$\tau_\mathrm{rot}$ {[ps]}\\\hline
MQ	&0.24 &0.56 &44\\
EQ	&0.27 &0.65 &52\\
iPrQ     &0.25&0.59 &59\\
3iPr-MQ   &0.25 &0.61 &64\\
3Am-MQ   &0.31&0.71 &65\\
3tBu-MQ   &0.34 &0.80 &64\\
 \hline\hline\\\
 \end{tabular}
 \label{TAB:taures} 
\end{table}

To complement the solvation dynamics analysis, we also calculated water dynamics at specific sites around the solute so that 
the origin of the effect of a hydrophobic side chain can be located.
To this aim we analyzed the site-specific (semi-atomic) residence times from nonequilibrium simulations
around the unchanged parts of the different oxyquinolinium betaine derivatives, similar to the analysis of residence times in Ref.~\citenum{sch16c}. 
Only the water molecules which resided at the instance of excitation in the solvation shell of one of the specified groups listed in Table~\ref{TAB:res} were monitored.
The definition of solvation shells was made via Voronoi tessellation~\cite{vor08a}. The trajectory was then unfolded for these 
molecules and their location analyzed, insofar as whether the molecule is still (or again) in the first shell of its initial neighboring group in the simulation box (not in one of its images).
Longer stays at the surface of the solute are  connected to retardation of solvent motion via changed solute-solvent interactions, geometric constrains or collective effects. 
\begin{table}[t]
 \centering
 \caption{Retardation $\tau_{\mathrm{res},\mathrm{solute}}/\tau_{\mathrm{res},\mathrm{MQ}}$ of site-specific residence times at \SI{27}{\celsius}, atom labeling as in Fig.~\ref{FIG:derivatives}.}
 \begin{tabular}{lC{0.9cm}C{0.9cm}C{0.9cm}ccc}
 \hline\hline\
Atoms   &MQ&EQ&iPrQ&3iPr-MQ&3Am-MQ&3tBu-MQ\\\hline
C$_2$+H$_2$	&1.00	&1.06	&1.14	&1.16	&1.11	&1.15\\
C$_4$+H$_4$	&1.00	&1.04	&1.12	&1.15	&1.10	&1.12\\
C$_\mathrm{4A}$&1.00	&1.04	&1.20	&1.22	&1.18	&1.18\\
C$_5$+H$_5$	&1.00	&1.02	&1.09	&1.17	&1.15	&1.15\\
C$_6$+O$_6$	&1.00	&1.02	&1.12	&1.12	&1.12	&1.09\\
C$_7$+H$_7$	&1.00	&1.07	&1.15	&1.12	&1.11	&1.08\\
C$_8$+H$_8$	&1.00	&1.06	&1.10	&1.13	&1.11	&1.09\\
C$_\mathrm{8A}$&1.00	&1.03	&1.16	&1.18	&1.16	&1.18\\
 \hline\hline\\\
 \end{tabular}
 \label{TAB:res} 
\end{table}
Table~\ref{TAB:res} lists the retardation of residence times of the different oxyquinolinium betaine derivatives. 
Residence around EQ is slightly retarded compared to MQ, whereas iPrQ, 3iPr-MQ, 3Am-MQ and 3tBu-MQ show stronger retardation of residence in the first shell. The retardation of residence times
is usually stronger at sites close to the introduced side chain and weaker (but still evident) at sites further away.
We attribute at least parts of the effect to geometric constraints, as a water molecule close to a bulky side chain has less directions to leave the chromophore surface, compared to a water molecule close to the unsubstituted chromophore. 
The overall retardation of residence is about 1.1 for the larger derivatives, indicating that there is a slight change in water dynamics, which contributes to the observed retardation of solvation dynamics.
However, it cannot account for the full retardation of $\tau_\mathrm{solv}$ of 1.3 to 1.4 of the experimental Stokes shift relaxation function, so that there have to be other contributions, too. 
For comparison, when trehalose, which is known to affect 
water dynamics greatly, was attached to the chromophore instead of a hydrophobic side chain, the retardation of site-specific residence times equaled the retardation of relaxation times and thus a change in water
dynamics in the trehalose system could be confirmed; see also Ref.~\citenum{sch16c}. For 3tBu-MQ, the contribution of changed water dynamics is  only marginal and the major source of retardation of solvation dynamics has to be sought elsewhere.

\begin{figure}[t]
  \includegraphics[width=\linewidth]{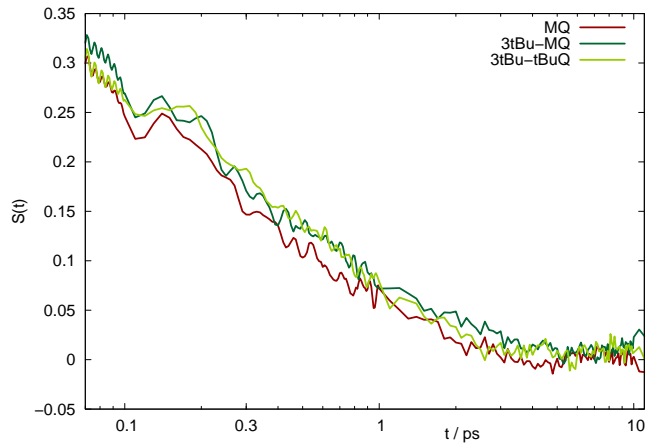}
 \caption{Stokes shift relaxation function of MQ, 3tBu-MQ and 3tBu-tBuQ at \SI{27}{\celsius}. The oscillations at times below \SI{0.1}{ps} correspond to fast water libration.}
 \label{FIG:large}
 \end{figure}

To verify the assumption that the observed retardation does not stem from hydrophobic interactions, we conducted a simulation of an even larger chromophore. For 3tBu-3tBu,  tertbutyl moieties where attached both to nitrogen
and to C$_3$ of the oxyquinolinium backbone. Upon observance of hydrophobic interactions, 
the relaxation time of this new compound is expected to be slower than for 3tBu-MQ.
The Stokes shift relaxation curves of 1MQ, 3tBu-MQ and 3tBu-3BuMQ are shown in Fig.~\ref{FIG:large}.
The introduction of larger side chains on the Stokes shift relaxation function showed no effect, so that the so far drawn assumptions are corroborated. 
Thus, other possible sources of retardation, namely solute rotation, translation and vibration, are examined in the following.

\subsection{Solute rotation times}
To detect a possible coupling of solute rotation and the solvation response, we calculated rotation times
of solute (and solvent) obtained from the second Legendre polynomial of the normalized
correlation function of the corresponding dipole vector from equilibrium simulations of the ground state solutes. 
The derivatives differ in their rotation time, as listed in Table~\ref{TAB:taures} and shown graphically in the Supporting Information.
The larger the hydrophobic side chain, the longer the solute needs to rotate. All of them rotate
significantly slower than water ($\tau_\mathrm{rot} \simeq 4\mathrm{ps}$).
As solvation dynamics comes mainly from rotation (and translation) of solvent molecules relative to the solute, solute rotation only takes effect if it occurs 
on the same or a faster timescale as the solvent rotates~\cite{ern11a}.
Water rotates very fast, so that the different rotation times of the oxyquinolinium betaine derivatives should not
affect the solvation relaxation times significantly. Therefore, the same solvation response is expected for all 
solutes in water.

\subsection{Restrained nonequilibrium simulations: Contribution of solute and solvent}
\begin{figure}[t]
  \includegraphics[width=\linewidth]{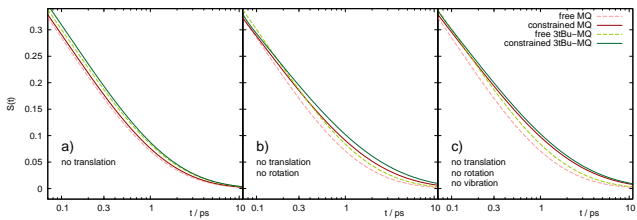}
  \caption{Restrained nonequilibrium simulations of MQ and 3tBu-MQ at \SI{20}{\celsius} (fitting functions). a) Omission of translation, b) omission of translation
    and rotation, c) omission of translation, rotation and vibration.}
 \label{FIG:restrain}
 \end{figure}

To investigate the absolute effect of solute translation, rotation and vibration, selected atom coordinates in MQ and 3tBu-MQ
were frozen after excitation. To omit translation, the coordinates of the atom C$_\mathrm{8A}$, which is close to the center of mass,
were not allowed to change during relaxation. This restraint has no influence, neither on the relaxation around MQ nor on 3tBu-MQ, as shown
in Fig.~\ref{FIG:restrain}a.
 To further omit rotation, the coordinates of atoms C$_\mathrm{8A}$, C$_3$ and C$_6$ were fixed. The respective curves are shown
 in  Fig.~\ref{FIG:restrain}b. The omission of rotation slightly slows down the response
 of both MQ and 3tBu-MQ alike, but changes the observed retardation only to a very small extent.
  Only when all degrees of freedom (translation,
 vibration and rotation) are omitted, 3tBu-MQ show nearly the same Stokes shift relaxation function, shown in Fig.~\ref{FIG:restrain}c. Here, the relaxation curve of 3tBu-MQ was only slightly affected (namely, retarded) by this restraint,
 whereas the solvent around MQ relaxed more slowly compared to the previous restraints or the unrestrained simulation.
 A tiny retardation, however, persists, which might be attributed to hydrophobic interactions, as also found
   in the analysis of residence times. This indicates that
 for the small MQ, solute vibration contributes to solvation dynamics. 
 For reasons of clarity, Fig.~\ref{FIG:restrain} shows only the fitting functions of the actual data points. The unprocessed data is shown in the Supporting Information, were the same trends are clearly visible.
 
 We note that the freezing of a set of three atoms to avoid rotation also affects some normal modes. We thus 
 conducted four different simulations where rotation was omitted by freezing different atom coordinates.
 The corresponding
 curves are shown in the Supporting Information. Whenever the frozen atoms are located mostly at the pyridinium part of the
 molecule, a larger slowing down of the dynamics around MQ is found, so that the difference between MQ and 3tBu-MQ becomes smaller or disappears. We assume that some selections affect the possible normal
 modes of MQ, and thus slow down solvation dynamics. The fast dynamics around MQ is therefore achieved by some normal modes
 located at the pyridinium part of the chromophore that contribute to relaxation. By addition of a sidechain, or selective
 freezing of atoms, these normal modes do not occur any more and the relaxation around MQ is slowed down, yielding (approximately) the slower response around 3tBu-MQ.
 The corresponding relaxation times of all restrained simulations and retardation factors are listed in Table~\ref{TAB:con}, where the small differences seen in Fig.~\ref{FIG:restrain} reproduce the retardation factors discussed above. Only
 the selective freezing of some normal modes (TR$_\mathrm{H_2,H_5,H_8}$ or TR$_\mathrm{N_1,C_3,C_{4A}}$), or of all degrees of freedom (TRV) leads to a significant change in retardation (for nomenclature, see Table~\ref{TAB:con}). 
 \begin{table*}[t]
 \centering
 \caption{Relaxation times in ps of free and constrained simulations, where combinations of translation (T), rotation (R) and vibration (V) were omitted.}
 \begin{tabular}{lrrrrrrr}
 \hline\hline\ 
		& free 	&T 	& TR	& TR 	&TR	& TR	&TRV\\
frozen atoms	&	&\tiny{$\mathrm{C_{8A}}$}&\tiny{$\mathrm{C_{8A},C_3,C_6}$}&\tiny{$\mathrm{H_2,H_5,H_8}$}&\tiny{$\mathrm{N_1,C_3,C_{4A}}$}&\tiny{$\mathrm{C_{4A},C_6,C_8}$}&\tiny{all}\\\hline
 MQ		& 0.30	&0.31			&0.47				&0.51				&0.47				&0.45				&0.53	\\
 3tBu-MQ 	&0.38	&0.39			&0.60				&0.47				&0.41				&0.63				&0.58	\\
 Retardation 	&1.3	&1.3			&1.3				&0.9				&0.9				&1.4				&1.1	\\
 \hline\hline\\\
 \end{tabular}
 \label{TAB:con} 
\end{table*}
We note, however, that these differences are very small, and that the partitioning to hydrophobic effects and solute motion is only qualitative, not quantitative.

We furthermore conducted simulations where the  water oxygen (which is nearly equivalent to the respective center of mass) was also fixed, in addition to a completely fixed solute, so that relaxation after solute excitation could occur solely via water rotation.
The relaxation occurs very fast (in about \SI{0.05}{ps}), and no difference could be detected between MQ and 3tBu-MQ. Corresponding data is shown in the Supporting Information. 
We of course neglect cross-terms between rotation and translation, 
which also contribute to the relaxation, so that such a restrained simulation does not depict a physical process. Nevertheless,
the early relaxation seems to consist partly of fast solvent rotation, which are unaffected by the presence of large sidechains.
 For comparison (and to validate this approach) we also analyzed the trehalose-water system from Ref.~\citenum{sch16c} using this procedure and found significant differences between
the Stokes shift relaxation functions from the restrained MQ and trehalose-MQ.
We therefore do not 
find a large influence of the hydrophobic substitution on solute-solvent interaction itself (\textit{e.g.} via changed solvent structure). Rather, the influence of solute vibration is different for MQ and 3tBu-MQ, so that different 
solvation dynamics are observed.

\subsection{Analysis of chromophore vibrations}
\begin{figure}[t]
  \includegraphics[width=\linewidth]{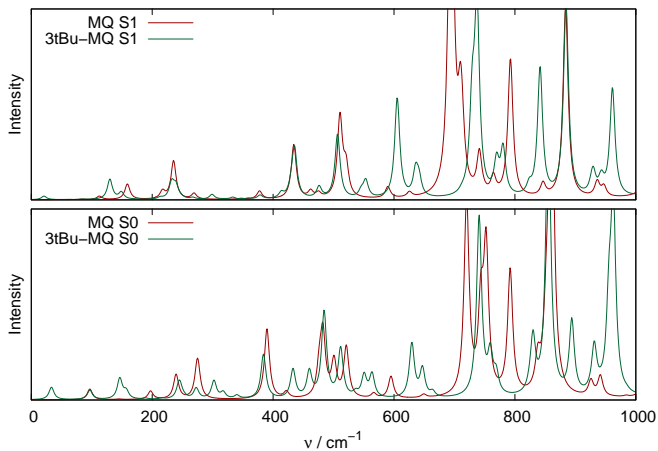}
 \caption{Normal mode spectra (IR) up to \SI{1000}{cm}$^{-1}$ of MQ and 3tBu-MQ in the ground (bottom) and excited (top) state.}
 \label{FIG:spectra}
 \end{figure}

 To see the changes in the low frequency modes upon addition of a sidechain we calculated vibrational spectra from quantum
mechanics (optimization and normal mode calculation at the B3LYP 6-311G++(2d,2p) level of theory) in the ground and excited state of MQ and 3tBu-MQ, shown in Fig.~\ref{FIG:spectra}. 
 Upon addition of the sidechain, the normal modes change to a large extent. 
  The same observations were made for the normal modes calculated from molecular dynamics simulation 
 using the force fields developed in this study. All normal modes from quantum mechanics and molecular dynamics simulation are listed in the Supporting Information.
 Interestingly, many of the low frequency modes involve movement of H$_2$, H$_3$ and H$_4$, which is hindered if H$_3$ is exchanged 
 to a larger sidechain. 
 The importance of these vibrational modes were already noticed when fixing atoms of the pyridinium part of the solute in the previous section (see also the Supporting Information).
 Thus, the tertiary butyl group may hinder the coupling of some of the modes to hydration water and thus slow down solvation dynamics.
  Together with the observation that 3tBu-tBuQ does not show slower solvation dynamics than 3tBu-MQ, and that selective freezing of solute atoms affects
 the degree of retardation between MQ and 3tBu-MQ, we can therefore conclude that solute vibration is mainly responsible for the retardation of oxyquinolinium betaine derivatives upon sidechain attachment.

\section{Conclusion}
We examined the molecularity of aqueous solvation dynamics via simulation of the time-dependent Stokes shift of different derivatives of oxyquinolinium betaine, differing only in location
and length of hydrophobic side chains. 
Such a dependence of solvation dynamics on solute size was so far only known for slowly rotating solvent as methanol, where solute and solvent rotate on the same timescale.
As  water rotates faster than any of the solutes studied, 
the change in rotation times of the different oxyquinolinium derivatives should be 
negligible and no solute dependence of aqueous solvation dynamics has been described yet (unless where changes in hydrogen bonding occur).
However, high resolution measurement published in the companion paper, Ref.~\citenum{ern17a} showed a small but non-negligible influence.
Likewise, we found that the solvation times increased with increasing chain length also in simulation.
The detection of these small differences
 were made possible only by current advances in technology (broadband upconversion) and simulation (increased computational power), as  formerly such small effects where lost in statistical noise.
The activation barrier of solvation for MQ was found to increase by
about \SI{5.4}{kJ/mol} upon inclusion of a tertiary butyl group. The water molecules around solutes containing
large aliphatic side chains therefore seemed to be hindered to relax to a change in electric field.

The analysis of semi-atomic residence times across
the solute showed that the attachment of side chains to the solute resulted in a retardation of first shell residence times. This retardation can be attributed to hydrophobic interactions and geometric effects, but accounts only for 
a small fraction of the observed overall retardation of solvation dynamics. The major part of the retardation was found to stem from solute motion for small derivatives, whereas for larger derivatives solute motion is negligible. 
In fact, in the excited state MQ and 3tBu-MQ interact differently with the surrounding solute, where the difference vanishes if the solute is made rigid. 
Here, the omission of translation does not affect the solvent response and the omission of rotation uniformly slows down solvation to a small extent around all chromophores alike. The further omission of vibration slows down the MQ response so
that no more retardation compared to the larger tertbutyl derivative can be found. We note that due to the smallness of the effect, such a assignment is only qualitative, not quantitative. 

The observed retardation in Ref.~\citenum{ern17a} therefore stems only to a small extent from hydrophobic interactions, but mostly from solute motion, presumably vibration that couples differently to the solvation response.  
The attachment of a hydrophobic group to a chromophore thus changes the overall solvent dynamics even for solvents like water.
The present picture of invariance of solvation dynamics to solute structure -- as long 
as the solvent rotates faster than the solute and the hydrophilic parts of the solute are not changed --  therefore needs to be revised.

\section{Associated Content}
The Supporting Information is available free of charge on the ACS Publications website.

Forcefields and geometries of all employed chromophores, 
discussion of the validity of linear response theory for oxyquinolinium betaines, 
solvation response for alternative freezing patterns for the omission of solute rotation,
visualization of solvent rotation contribution, 
correlation of rotation and solvation times,
influence of the partial charge distribution of 3tBu-MQ, 
influence of geometry adaption on the time-dependent Stokes shift,
normal mode frequencies of MQ and 3tBu-MQ from quantum mechanism and molecular dynamics simulation

\section{Acknowledgement}
We would like to thank M. Gerecke and N. P Ernsting for fruitful discussions and communication of their experimental results prior to publication.
This work was funded by the Austrian Science Fund FWF in the context of Project 
No. FWF-P28556-N34. E.H. is  recipient of a DOC Fellowship of the Austrian Academy of Sciences at the Institute of Computational Biological Chemistry.
The computational results presented have been achieved in part using the Vienna Scientific Cluster (VSC).
\providecommand{\latin}[1]{#1}
\providecommand*\mcitethebibliography{\thebibliography}
\csname @ifundefined\endcsname{endmcitethebibliography}
  {\let\endmcitethebibliography\endthebibliography}{}

\end{document}